\documentclass{ws-ijmpa}

\begin{document}

\markboth{A. Doff and A. A. Natale}
{A dynamical mechanism  to explain the top-bottom quark mass hierarchy}

%
\catchline{}{}{}{}{}
%

\title{A dynamical mechanism  to explain the top-bottom quark mass hierarchy}

\author{A. Doff  and A. A. Natale}

\address{Instituto de F\'{\i}sica Te\'orica,
Universidade Estadual Paulista,
Rua Pamplona 145,
01405-900, S\~ao Paulo, SP,
Brazil\\
doff@ift.unesp.br \\
natale@ift.unesp.br}

\maketitle


\begin{abstract}
We discuss the mass splitting between the the top and bottom quarks in a  technicolor scenario.
The model proposed here contains a left-right electroweak gauge group. An extended technicolor group and  mirror fermions  are introduced. The top-bottom quark mass splitting turns out to be intimately  connected to the breaking of the left-right gauge symmetry. Weak isospin violation occurs within the experimental limits.
\end{abstract}

\ccode{PACS numbers: 12.60.Nz, 12.10.Dm, 14.80.Cp}

\def\br{\begin{eqnarray}}
\def\er{\end{eqnarray}}
\def\be{\begin{equation}}
\def\ee{\end{equation}}
\def\a{\alpha}
\def\D{\Delta}
\def\d{\delta}
\def\g{\gamma}
\def\G{\Gamma}
\def\l{\lambda}
\def\L{\Lambda}
\def\m{\mu}
\def\n{\nu}
\def\({\left(}
\def\){\right)}
\def\<{\left\langle}
\def\>{\right\rangle}
\def\s{\sigma}
\def\S{\Sigma}
\def\cao{\c c\~ao\,}

\section{Introduction}

In the standard model of elementary particles the fermion and gauge boson
masses are generated due to the interaction of these particles with elementary
Higgs scalar bosons. Despite the success of the model there are some intriguing points
as, for instance, the enormous  range of masses between the lightest and heaviest
fermionic generations. In particular we can also quote the large ratio between the Yukawa
coupling constants $Y_{{b}}/Y_{{t}} \sim  1/41$, required to give  the correct masses to the
top and bottom quarks. These questions  could be better explained with the introduction of new fields and
symmetries. One of the possibilities in this
direction is the substitution of elementary Higgs bosons by composite ones in the
scheme named technicolor \cite{hs,lane}.

The beautiful characteristics of technicolor (TC) as well as its problems
were clearly listed recently by Lane \cite{lane}. Most of the technicolor
problems may be related to the dynamics of the theory as described in
Ref.\cite{lane}. Although technicolor is a non-Abelian gauge theory, it is
not necessarily similar to QCD, and if we cannot even say that QCD is fully understood
up to now, it is perfectly reasonable to realize the enormous work that is needed
to abstract from the fermionic spectrum the underlying technicolor dynamics.

The many attempts to build a realistic model of dynamically generated
fermion masses are reviewed in Ref.\cite{hs,lane}. Most of the work in this area
try to find the TC dynamics dealing with the particle content of the
theory in order to obtain a technifermion self-energy that does not
lead to phenomenological problems,  as an example we can consider the scheme known as walking
technicolor \cite{walk}. More recent papers envisage such different dynamics as a consequence of a conformal symmetry\cite{luty}. In this work  we propose a model to explain the
large (t-b) splitting in the context of a technicolor scenario.

In  order to obtain  R $\sim \frac{1}{40}$, where R $= \frac{m_{b}}{m_{t}}$,  we  will consider a
technicolor model based on the gauge group
\br {\sl G} = SU(3)_{{}_{ETC}} \otimes G'
\er
where in the above structure  we define $G'= SU(3)_{{}_{C}}
\otimes SU(2)_{{}_{L}}\otimes SU(2)_{{}_{R}} \otimes U(1)_{{}_{S}}$. In  this model we consider that the gauge group $G'$ is embedded in  a grand-unified theory (GUT) with a gauge group $SO(12)$ \cite{rajpoot}. This group contains the electroweak Left-Right model that has been widely studied\cite{LR} and naturally contains a complete fermionic  generation ($\Psi$) as well as  its respective mirror partner ($\widehat{\Psi}$). An explanation of why we have performed this particular choice will become clear in the fourth section. Models along this line have been proposed by Appelquist and Schrock\cite{schrock}. The variation  in our case is the introduction of the  mirror sector.  An interesting feature of our model is that  we are able of  the relate the top-bottom quark  mass splitting  as ratio between Left-Right boson masses.
The  fermionic content of the model is divided into two distinct sectors, the sector $(a)$ containing $t$ and $b$ quarks, $T$ and $B$ techniquarks and their respective leptons. The sector $(b)$, or mirror sector\,($\widehat{{\,\,\,}}$), contains the mirror partners $\hat{t}$ and $\hat{b}$  with  respective  mirror leptons.

\par In the above model we do not introduce an ETC gauge sector associated to the mirror fermions because  the mass scale  peculiar to these particles is expected to be around (1-10)TeV\cite{maalampi}. Therefore in this scenario it can be natural to consider  that the mirror particles obtain their masses through the same mechanism responsible for the $SO(12)\rightarrow SU(3)_{{}_{C}}\otimes SU(2)_{{}_{L}}\otimes  U(1)_{{}_{Y}}$ symmetry breaking. In this model only the usual fermions and LH gauge bosons will receive their masses dynamically. The mirror fermions and the RH gauge bosons will be very heavy, and their masses could result from the existence of elementary scalar fields which can naturally appear at the GUT scale.

To ensure that  $m_{t}\neq m_{b}$ without introducing large violations of weak isospin we will assume an scheme proposed by Sikivie et al. some years ago \cite{sikivie}. The basic idea of this mechanism is to put
the chiral components (LH and RH) of a given fermion, member of an electroweak doublet,  in  different
ETC representations. For example, in our model the chiral components of the bottom quark will be in
different ETC representations, therefore the bottom quark does not receive mass at leading order as
happens for the top quark.  However in our model diagrams as the ones depicted in the Fig.(1) will contribute  at higher order to the bottom quark mass.

\section{The fermionic content of the model}

According to the group structure ${\sl G}$ and following the recipe of Ref.\cite{sikivie} we can now choose
the fermionic representation in such a way that the bottom quark mass arise from diagrams of the type shown in Fig.(1). This point will be  discussed with more details in the section 4.

The fermions and mirror fermions  are assigned to  the following representations of the gauge group
$SO(12)$ in terms of the $SO(10)$ group decomposition
\\
\noindent(a) The fermion sector
\br
{\bf{16}}^T  =  ( t_{1}...t_{3}\, ,\nu_{\tau}, b_{1}... b_{3} \,, \tau^{-} \,, b^{c}_{1}...b^{c}_{3}\,,  \tau^{+} \,, -t^{c}_{1}... -t^{c}_{3}\,, -\nu^{c}_{\tau})_{L}
\er
\noindent (b) The mirror sector
\br
{\bf{16}^{*}}^T  =( \hat{t}_{1}...\hat{t}_{3}\, , \hat{\nu}_{\tau}\, , \hat{b}_{1}... \hat{b}_{3}\,,   \hat{\tau}^{-}\, , {\hat{b}}^{c}_{1}... {\hat{b}}^{c}_{3}\,, \hat{\tau}^{+} \,, -{\hat{t}}^{c}_{1}...  -{\hat{t}}^{c}_{3} \,, -{\hat{\nu}}^{c}_{\tau})_{L}
\er
\noindent where the above structure form an irreducible spinor $\Psi^{+} = (\Psi^{{\bf 16}} \,,\, \widehat{\Psi}^{{\bf 16^{*}}})$ of the $SO(12)$, and the index $i = 1..3$ is a color index.
The $SO(N)$ gauge theories are naturally free  from anomalies,  the anomalies involving the ETC sector are  canceled by introduction of the respective leptons and technileptons. The mirror fermions in our model are introduced in order to implement the Barr and Zee mechanism\cite{zee}, which has been used to explain the muon-electron mass difference. In our case the mirror fermions will replace the exotic leptons present in the Barr and Zee model.

The  fermionic  assignments  of the ${\bf{16}}^T$ multiplet shown in  Eq.(2)  with respect to the  sub-group $SU(3)_{{}_{ETC}}\otimes SU(3)_{{}_{C}}\otimes SU(2)_{{}_{L}}\otimes SU(2)_{{}_{R}} \otimes U(1)_{{}_{S}}$ are the following
\begin{eqnarray}
&&t^i_{{}_{L}} \,\sim \,(3,3,2,1,1/3)\,\,\,,\,\,\,t^i_{{}_{R}} \,\sim \,(3,3,1,2,1/3)\nonumber \\
&&b^i_{{}_{L}} \,\sim \,(3,3,2,1,1/3)\,\,\,,\,\,\,b^i_{{}_{R}} \,\sim \,(\bar{3},3,1,2,1/3)\nonumber \\
&&\tau_{{}_{L}}\,\sim \,(3,1,2,1, -1)\,\,\,,\,\,\,\tau_{{}_{R}}\,\sim \,(3,1,1,2, -1)\nonumber\\
&&\nu_{\tau_{{}_{L}}}\,\sim \,(3,1,2,1, -1)\,\,\,,\,\,\,\nu_{\tau_{{}_{R}}}\,\sim \,(\bar{3},1,1,2,-1).
\nonumber
\end{eqnarray}

The mirror fermionic assignments are similar, however mirror fermions  are singlets of $SU(3)_{{}_{ETC}}$, as  discussed in the introduction. The technifermions quantum numbers are the same ones as the ordinary fermions. The only difference between fermions and technifermions are the TC quantum numbers. Besides that $b$ and $t$ are in different ETC representations.

In  theories where the  electroweak interaction is broken dynamically,  the  fermionic
mass terms  arise   when  the ETC gauge group is broken \cite{ETC}. We shall not  speculate about the origin of this breaking. However, we will suppose that this interaction and the $SO(12)$ gauge symmetries  can be broken by fundamental scalars associated to supersymmetry or any other mechanism at the GUT scale. In the reference\cite{rajpoot} the breaking of this GUT to $SU(3)_{{}_{C}}\otimes SU(2)_{{}_{L}}\otimes  U(1)_{{}_{Y}}$ is discussed in detail. For convenience we will reproduce the symmetry breaking pattern of this model
\br
SO(12)&&\stackrel{M_{{}_{a}}}{\longrightarrow}SU(4)_{{}_{F}}\otimes SU(4)_{{}_{C}} \nonumber \\ &&\stackrel{M_{{}_{b}}}{\longrightarrow}SU(4)_{{}_{C}} \otimes SU(2)_{{}_{L}}\otimes SU(2)_{{}_{R}} \otimes U(1)_{{}_{F}} \nonumber \\
&&\stackrel{M_{{}_{c}}}{\longrightarrow}SU(3)_{{}_{C}} \otimes SU(2)_{{}_{L}}\otimes SU(2)_{{}_{R}} \otimes U(1)_{{}_{C + F}} \nonumber \\
&&\stackrel{M_{{}_{d}}}{\longrightarrow}SU(3)_{{}_{C}} \otimes SU(2)_{{}_{L}}\otimes SU(2)_{{}_{R}} \otimes U(1)_{{}_{S}} \nonumber \\
&&\stackrel{M_{{}_{R}}}{\longrightarrow}SU(3)_{{}_{C}} \otimes SU(2)_{{}_{L}} \otimes U(1)_{{}_{Y}}.
\label{eq3}
\er
\begin{figure}[t]
\begin{center}
\epsfig{file=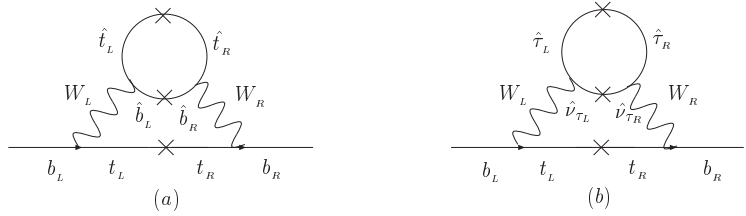,width=0.8\textwidth}
\caption{Diagrams contributing to  the bottom  quark mass.
In the diagram (a) the contribution to the $b$ quark mass comes from the mirror-quarks  of our model while the contribution of the diagram (b) is due to the mirror-leptons.}
\end{center}
\end{figure}
\!where  we indicate $U(1)_{{}_{C + F}} = U(1)_{{}_{C}}\otimes U(1)_{{}_{F}}$ .
The masses scales ($M_{a..d}$, $M_{{}_{R}}$)  related to the symmetry breaking of $SO(12)$ group down to $SU(3)_{{}_{C}} \otimes SU(2)_{{}_{L}} \otimes U(1)_{Y}$  are determined respecting the constraints on the proton decay lifetime and the renormalization of the weak angle bare value $\sin^2\theta_{{}_{W}}$ to the experimental value $\sin^2\theta_{{}_{W}}=0.23120(15)$\cite{PDG}.  In our model the symmetry breaking of the last stage shown in Eq.(\ref{eq3}) will be implemented by a techniquark condensate characterized by a strong interaction $SU(2)_{{}_{TC}}$ symmetry at $O(300)GeV$. According  to what we discussed above we assume that the breaking of the $SU(3)_{{}_{ETC}} \rightarrow SU(2)_{{}_{TC}}$ will happens  near  the GUT scale.

We shall  concentrate only  in the low energy sector (i.e. bellow $1$ TeV) of our model, $SU(3)_{{}_{C}}\otimes SU(2)_{{}_{L}}\otimes SU(2)_{{}_{TC}}\otimes U(1)_{{}_{Y}}$, and in particular in the top-bottom mass splitting without touching the problem of mass generation for the remaining fermions.

\section{The Top Dynamical Mass}

We  can decompose the fermionic content associated to the gauge group $SU(3)_{{}_{ETC}}$ as shown in the following:

\br
&& \psi_{{}_{{{\bf{3}}}_{ETC}}} = \left(\begin{array}{c}   T_{1,\omega} \\ T_{2,\omega} \\ t \end{array}\right)_{{L,R}}\,,\,\,\, \,\,\,\,\left(\begin{array}{c}   B_{1,\omega} \\ B_{2,\omega} \\ b \end{array}\right)_{L}\,\,\,\sim \,\,\,\,\,({\bf{3}}, 1) \nonumber \\
&& \psi_{{}_{\bar{\bf{3}}_{ETC}}} = \left(\begin{array}{c}  B_{1,\omega} \\ B_{2,\omega} \\ b \end{array}\right)_{R}\,\,\,\sim \,\,\,\,\,({\bar{\bf{3}}}, 1)
\label{eqdecom}
\er
\noindent  where  $\omega$ is a techniflavor index indicating the number($n_{\omega}$)  of flavors necessary to produce  the walking behavior for the TC group. In Eq.(\ref{eqdecom}) we indicate only the $SU(3)_{{}_{ETC}}$ decomposition, the $SU(2)_{L}$ structure of $(t\,,\,b)_{{}_{L}}$ multiplet  is the usual one.
\par This choice  is in agreement with the scheme proposed by Sikivie et al. \cite{sikivie} in order to assure  that $ m_{b} \neq m_{t}$ without large weak isospin violation, where we are putting  the (LH - RH) chiral components of the top quark  in the same ETC representation, but the bottom quark components are in different ETC representations. In this way the dynamical top quark mass is generated at leading order as depicted in the Fig.(2), as  happens  in ordinary technicolor models.

\begin{figure}[t]
\begin{center}
\epsfig{file=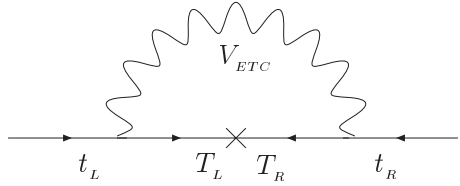,width=0.5\textwidth}
\caption{Diagram responsible for the top quark mass.}
\end{center}
\end{figure}

The top quark mass is giving by the following expression:

\be
m_{t} \approx \frac{3g^2_{{}_{ETC}}C_{{}_{ETC}}}{8\pi^2}
\int^{\Lambda_{{}_{ETC}}}\!\!\!\frac{d^4p}{(p^2 + \Lambda^2_{{}_{TC}} ) }\frac{\Sigma(p)_{{}_{TC}}}{(p^2 + \Lambda^2_{{}_{ETC}})} ,
\label{masst}
\ee
while at this order we have $m_{b} = 0,$ and $\Lambda_{{}_{ETC}}$ is the mass scale of the  ETC bosons.
It is useful  to  write  the  technifermion self-energy in terms of the ansatz formulated in the Ref. \cite{aa1}
\be
\frac{\S(p)_{{}_{TC}}}{ \Lambda_{{}_{TC}}} = \(\frac{\Lambda^2_{{}_{TC}}}{p^2}\)^{\a}\!\!\!\![1 + bg^2(\Lambda^2_{{}_{TC}})ln(p^2/\Lambda^2_{{}_{TC}})]^{-\gamma_{{}_{TC}}{cos(\a\pi)}}
\label{stec}
\ee
\noindent where, in this  equation,  we identify $\gamma_{{}_{TC}} = \frac{3C_{{}_{TC}}}{16\pi^2b}$ with $C_{{}_{TC}}$ given by
$C_{{}_{TC}} =\frac{1}{2}[C_{2}(R_{1}) + C_{2}(R_{2}) - C_{2}(R_{3})]$,
where $C_{2}(R_{i})$ (with $i = 1..3$) is the Casimir operator for the  technifermions in the representations $R_{1}$ and $R_{2}$
that condensate in the representation $R_{3}$, and $b$ is  the coefficient of order $g^3$ in the technicolor $\beta_{{}_{TC}}$ function. The quadratic Casimir operator  for the $ETC$ gauge group can be written as $C_{{}_{ETC}} = (N^2_{{}_{ETC}} - 1)/2N_{{}_{ETC}} = 4/3$ for $N_{{}_{ETC}}= 3$. Finally, the mass scale $\Lambda_{{}_{TC}}$ can be estimated to be of the order of the electroweak symmetry scale $\Lambda_{{}_{TC}} \sim O(300)\,GeV$, according the discussion performed in the last section the scale $\Lambda_{{}_{ETC}}$ can be assumed as $\Lambda_{{}_{ETC}} \approx \Lambda_{{}_{GUT}}$.

 Eq.(\ref{stec}) interpolates between  the standard operator product expansion result for the technifermion self-energy, which is  obtained when $\alpha \rightarrow 1$, and the extreme walking technicolor solution obtained when $\alpha \rightarrow 0$ \cite{walk}.
As we have pointed out in Ref. \cite{aa2} only such kind of solution  is naturally  capable of generating a large mass to the third fermionic generation, which has a mass limit almost saturated by the top quark mass.  Moreover, as also claimed in the second paper of Ref.\cite{aa2} there are other possible reasons to have $\alpha \sim 0$, as the existence of an IR fixed point and a gluon (or technigluon) mass scale \cite{alkofer}, which are closely related \cite{coup}. 

We will just consider $\alpha=0$ in Eq.(\ref{stec}) assuming that we can introduce a number of technifermions  necessary to obtain the extreme walking behavior,  our result for the t quark  mass (and consequently for b quark mass) will be dicussed exactly in this case.
Therefore, after  its substitution into Eq.(\ref{masst}), with  $p^2 = (\Lambda^2_{{}_{ETC}}/\Lambda^2_{{}_{TC}})y$, we verify that Eq.(\ref{masst}) is reduced to
\be
m_{t} \approx \frac{3g^2_{{}_{ETC}}C_{{}_{ETC}}\Lambda_{{}_{TC}} }{16\pi^2}\left[\frac{\!}{\!}1 + bg^2(\Lambda^2_{{}_{TC}})ln(\frac{\Lambda^2_{{}_{ETC}}}{\Lambda^2_{{}_{TC}}}) \right]^{-\gamma_{{}_{TC}}}\!\!\!\!\!\!\!\!\!I(\Lambda^2_{{}_{TC}})
\label{masst1}
\ee
where we used the definition
\be
I(\Lambda^2_{{}_{TC}})=\frac{1}{\Gamma(\delta)}\int^{\infty}_{0}\!\!\!\! dz e^{-z}z^{\delta - 1}\left(\frac{1}{\Lambda^2_{{}_{TC}}}\right)^{-\epsilon z}\!\!\!\int^{\Lambda^2_{{}_{TC}}}\!\!\frac{dyy^{-\epsilon z}}{y + \Lambda^2_{{}_{TC}}}.
\label{masst2}
\ee
To obtain the above equation we made use of the following Mellin transform:
\be
\left[1 + \epsilon\ln\frac{A}{B}\right]^{-\delta} = \frac{1}{\Gamma(\delta)}\int^{\infty}_{0}dzz^{\delta -1}e^{-z}\left(\frac{A}{B}\right)^{-\epsilon z}
\ee
and  defined $\epsilon = bg^2(\Lambda^2_{{}_{TC}})$, with  $\delta = \gamma_{{}_{TC}} + 1$.
After integration of the Eq.(\ref{masst2}) we can show that  Eq.(\ref{masst1}) leads to \cite{aa2}
\br
m_{t} \sim C_{{}_{ETC}}\alpha_{{}_{ETC}}\Lambda_{{}_{TC}}.
\er
The above result, which can naturally describe the top quark mass \cite{aa2}, resembles the way the top quark acquires a mass from an elementary Higgs boson with vacuum expectation
value  $\Lambda_{{}_{TC}}$. This behavior could be predicted based on the results of Ref.\cite{soni}, where it is remarked that when $\alpha \sim 0$ in Eq.(\ref{stec}) the composite Higgs system behaves as a fundamental one, and in this case some results obtained with the introduction of fundamental Higgs bosons are reproduced in a dynamical scenario.  This fact underlies several aspects of our model.

\section{The bottom  mass}

\par  According to what we discussed at the beginning of this work, the bottom  quark mass  will be generated  at 2-loop level, due to the interaction with mirror-quarks(leptons) and the  $W_{L,R}$ bosons as depicted in Fig.(1). There are other possible contributions to the $b$ quark mass, but they involve the  very massive $SO(12)$  gauge bosons and are negligible  compared to the ones of Fig(1). Assuming the usual Feynman rules,  we can write the following expression for the $b$ mass
\br
m_{b} = -i\frac{\alpha_{{}_{L}}\alpha_{{}_{R}}}{\pi^2}\int \frac{d^4q}{(q^2 - m^2_{t})}\left[\Gamma^{\mu}D_{\mu\beta}(p-q)I^{\beta\delta}(l,p,q)D_{\delta\nu}(p-q)(\not{\!q} + m_{t})\Gamma^{\nu}\right] \nonumber \\
\er
where in the above equation we defined the function $I^{\beta\alpha}$ as the fermion loop integral
\be
I^{\beta\delta}(l,p,q) = i\int\frac{d^4l}{(2\pi)^4}Tr\left[\Gamma^{\beta}S_{F}(l + p - q)\Gamma^{\delta}S_{F}(l)\right].
\label{loop}
\ee
In these expressions we are introducing the following notation: $\Gamma^{\mu..\nu} = \gamma^{\mu..\nu}\sigma^{a..d}$. The $\sigma^{a..d}$, for $a..d =1,2$ or $3$,  are $SU(2)_{{}_{L,R}}$ generators and the $D_{\mu\nu}(p-q)$ is the $W_{{}_{L,R}}$ propagator  in the Landau gauge.  Eq.(\ref{loop}), after angular  and $l$ integration, can be written as
\br
\frac{m_{b}}{ m_{t}} \!\approx - f(\alpha,n_{{}_{\widehat{F}}})\!\int\!\!\frac{dq^2}{(q^2 + m^2_{t})}\frac{(q^2 + M^2_{\hat{t}})}{(q^2 + M^2_{{}_{L}})}\frac{q^2}{(q^2 + M^2_{{}_{R}})}.
\label{eqR}
\er
\noindent where $M^2_{\hat{t}} = 3m^2_{\hat{t}}/4$. In order to obtain  this last equation, for simplicity, we considered the  zero momentum approximation  for the external propagators, we neglected the momentum dependence of the fermion  self-energies inside the
loop and assumed only the first diagram shown in the Fig.(1).
The masses  $m_{{}_t}$ and $m_{{}_{\hat{t}}}$ are respectively the top and mirror-top masses, $ M_{{}_{L}}$ and $M_{{}_{R}}$ are the weak vector boson masses and  we introduced the function
\be
 f(\alpha,n_{{}_{\widehat{F}}}) = \frac{27}{64\pi^2}\alpha_{{}_{L}}\alpha_{{}_{R}}n_{{}_{\widehat{F}}}C^L_{{2}_F}C^R_{{2}_F},
\ee
\noindent  with $n_{{}_{\widehat{F}}}$ indicating  the  number of  mirrorquarks(leptons) and $C^L_{{2}_F} = C^R_{{2}_F} = 3/4$ . Notice that if we had considered the momentum
dependence in the self-energies of Fig.(1) we would be lost in factors that do not bring any useful information to our calculation. Moreover, as we argue in the sequence, even in
this approximation the calculation is finite.

Simple power counting shows that the integral in Eq.\-(\ref{eqR}) is {\sl apparently}
logarithmically divergent,  we stress two points: First, if we had
considered the momentum dependence in the self-energies of Fig.(1) the integral
would be softened. Second, and as well as important, the model that we are proposing  here bears some similarity with the one proposed  by Barr and Zee some years ago \cite{zee},
whose purpose was the calculation of the electron mass in terms
of the  muon mass in the case of elementary scalar bosons. In that model a ``similar" type of logarithmic divergence  also occurs, which was canceled among different contributions to the electron  mass; the
muon contribution and the contribution of a heavy lepton X. This cancellation occurs when  it is assumed that the muon mixes with this lepton X. In our model we can suppose that the top and mirror top  mixes at the $SO(12)$ scale, and the cancellation now will occur between the fermion and mirror fermion sector keeping the $b$ mass finite. In terms of this mixing we can write Eq.(\ref{eqR}) in the following form
\br
m_{{}_{b}} \approx &&- m_{t}(\omega)f(\alpha,n_{{}_{\widehat{F}}})\int\!\!\frac{dq^2}{(q^2 + m^2_{t})}\frac{(q^2 + M^2_{\hat{t}})}{(q^2 + M^2_{{}_{L}})}\frac{q^2}{(q^2 + M^2_{{}_{R}})} +  \nonumber \\
 &&- m_{\hat{t}}(\omega)f(\alpha,n_{{}_{\widehat{F}}})\int\!\!\frac{dq^2}{(q^2 + m^2_{\hat{t}})}\frac{(q^2 + M^2_{\hat{t}})}{(q^2 + M^2_{{}_{L}})}\frac{q^2}{(q^2 + M^2_{{}_{R}})}.
\label{eqRM}
\er
where, for simplicity, we defined
\br
&&m_{t}(\omega) = m_{t}(\cos\omega_{{}_{L}})(\cos\omega_{{}_{R}}) \nonumber \\
&&m_{\hat{t}}(\omega)= m_{\hat{t}}(\sin\omega_{{}_{L}})(\sin\omega_{{}_{R}}).
\er
\par The relation between $m_{t}$ , $m_{\hat{t}}$  and the mixing angles $\omega{{}_{L,R}}$ as in the case of the Barr and Zee model\cite{zee} can be parametrized  as $m_{t}(\cos\omega_{{}_{L}})(\cos\omega_{{}_{R}} )=  -m_{\hat{t}}(\sin\omega_{{}_{L}})(\sin\omega_{{}_{R}})$. Then, after the integration of the  Eq.(\ref{eqRM}), we  can write the divergent pieces of this equation as
\br
m_{{}_{b}}(\Lambda) \propto  && m_{t}(\cos\omega_{{}_{L}})(\cos\omega_{{}_{R}})\log\Lambda^2  -  m_{t}(\cos\omega_{{}_{L}})(\cos\omega_{{}_{R}})\log\Lambda^2\!\!.
\er
Now we can assume that  $\Lambda\rightarrow \infty$  because  the combination of the two terms shown above give the finite result $m_{{}_{b}}(\Lambda)= 0$.

In the model of reference\cite{umemura}, which has also a common structure to the model discussed here, the authors estimated the mixing angle of the ordinary fermions with their mirror partners as being $\omega \leq 10^{-3}$, meaning that its cosine can be approximated by $1$. Therefore  our model has a larger convergence than the one in the Barr and Zee model due to the points discussed above.
Now we can justify our choice  for $SO(12)$ in order to unify the interactions at the scale $M_{{}_{GUT}}$, because this gauge group exactly accommodates one usual  fermionic generation and one mirror fermion generation necessary to keep the bottom mass finite in the present approach. This happens independently of the fact that we neglected the  momentum dependence of the self-energies in the loops of Fig.(1).

Finally, we can evaluate the contributions to the bottom quark mass shown in  Fig.(1). Diagram $(b)$ in Fig.(1) gives a contribution
identical to the diagram $(a)$, and the second can be obtained  from the first one just exchanging
in the calculation the mass $m_{{\hat{t}}}$  by  the mirror-lepton masses $m_{{\hat{\tau}}}$.
Taking into account both contributions we can write the ratio between the bottom and top quark masses as
\br
\frac{m_{b}}{ m_{t}} \approx  \frac{243}{1024\pi^2}\alpha_{{}_{L}}\alpha_{{}_{R}}n_{{}_{\widehat{F}}}\frac{M^2_{{}_R}}{(M^2_{{}_R} - M^2_{{}_L})}\left[\frac{(M^2_{{\hat{t}}} - M^2_{{}_R})}{(m^2_{t} - M^2_{{}_R})} + \frac{( M^2_{{\hat{\tau}}} - M^2_{{}_R})}{(m^2_{t} - M^2_{{}_R})} \right]\log\frac{M^2_{{}_R}}{M^2_{{}_L}}
\label{eq14}
\er
where we have already set $\cos\omega_{{}_{L,R}} \approx 1$.

It is important to note that in  this model  the  mass scale $M_{{}_{R(L)}}$  will be linked with the symmetry breaking scale of the $SO(12)$ group  by the following relations\cite{rajpoot}
\be
1 -  \frac{8}{3}\sin^2\theta_{{}_{W}} = \frac{11\alpha}{6\pi}\left[\frac{4}{3}\log\frac{M^2_{{}_{GUT}}}{M^2_{{}_L}} + 2\log\frac{M^2_{{}_R}}{M^2_{{}_L}}\right]
\ee
and
\be
\frac{\alpha}{\alpha_{s}} - \sin^2\theta_{{}_{W}} = \frac{11\alpha}{6\pi}\left[\log\frac{M^2_{{}_{R}}}{M^2_{{}_{GUT}}} -  \log\frac{M^2_{{}_R}}{M^2_{{}_L}}\right].
\ee

\noindent In order to keep $\sin^2\theta_{{}_{W}} = 0.231$  and $M^2_{{}_{GUT}} \propto 10^{16}GeV$  the symmetry breaking connected to the scale $M_{{}_{R}}$ must happens at $M_{{}_{R}} \sim 10^{13}GeV$. Therefore, we can  disregard the fermionic(mirror-fermionic) masses in comparison with the $M_{{}_{R}}$  mass and put Eq.(\ref{eq14}) into the following form

\be
\frac{m_{b}}{ m_{t}} \approx  \frac{243}{512\pi^2}\alpha_{{}_{L}}\alpha_{{}_{R}}n_{{}_{\widehat{F}}}\frac{M^2_{{}_R}}{(M^2_{{}_R} - M^2_{{}_L})}\log\frac{M^2_{{}_R}}{M^2_{{}_L}}.
\label{eq15}
\ee
\noindent This last equation is independent of the mirror fermions masses introduced in the model and as the Right-Handed symmetry is broken near at GUT scale the contribution associated to this  very high energy scale does not contribute to the $\rho$ parameter. We can regard this contribution as $\delta\rho \propto M^2_{{}_{L}}/M^2_{{}_{R}}$, therefore  we do not have too large  corrections to the Peskin-Takeuchi $T$ parameter, which is giving by \cite{pt}
\be
\alpha T = \frac{M^2_{W}}{M^2_{Z}\cos^2\theta_{W}} - 1 = \rho - 1 = \delta\rho.
\ee

In the figure below we plot the behavior of the bottom quark mass as a function of the ratio between the $M_{{}_{R}}$ and $M_{{}_{L}}$ masses.
\begin{figure}[ht]
\begin{center}
\epsfig{file=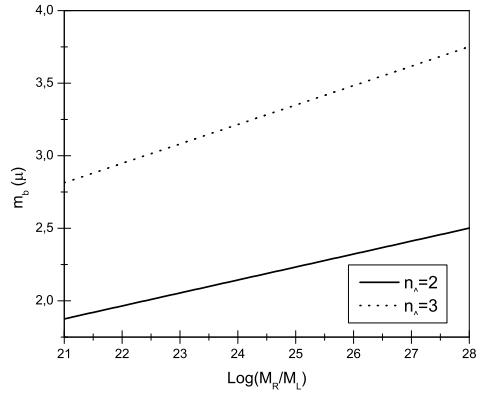,width=0.6\textwidth}\vspace{0.01cm}
\caption{The behavior of the bottom quark mass, $m_{b}(\mu)$, as a function of
${\frac{M^2_{{}_R}}{M^2_{{}_L}}}$. In the box we are indicating the number of
mirror flavors   $n_{{\,\,}_{\widehat{\,}}} \equiv n_{{}_{\widehat{F}}}= 2 $ and $n_{{}_{\widehat{F}}}= 3$ that we consider in our model. Only the strong radiative corrections have been considered to obtain these
curves.}
\end{center}
\end{figure}

The values of Fig.(2) correspond to the bottom quark mass at the scale $\mu^2 = 10\,GeV^2$, assuming that the bottom quark receives strong radiative corrections at low energies according to the correction factor given by \cite{buras}
\be
m_{b}(\mu) \approx m_{b}(\Lambda_{{}_{GUT}})\left(\frac{\alpha_{{}_{QCD}}}{\alpha_{{}_{GUT}}}\right)^{4/(11-\frac{2}{3}n_{{}_{f}})}.
\ee
\noindent For simplicity we considered only the $QCD$ radiative corrections  and we
also assumed $\alpha_{{}_{L}} = \sqrt{2}G_{{}_{F}}M^2_{{}_{L}}/\pi \sim 1/21$, $\alpha_{{}_{R}} \sim 1/37$\footnote{This value is obtained by extrapolation of the running coupling $\alpha_{{}_{R}}$ up to the scale $M_{{}_{R}} \sim 10^{13}GeV.$}, $\alpha_{{}_{GUT}} \sim 1/45$, $\alpha_{{}_{QCD}}(\mu^2 = 10 GeV^2)\approx 0.32$ with $m_{t}(\mu \approx 2M_{{}_{Z}}) \approx 178\,GeV$   and  $n_{{}_{f}} = 6$.  $G_{{}_{F}}$ and $\alpha_{{}_{QCD}}$ are respectively the Fermi coupling constant and QCD coupling constant.

\section{Conclusions}

 In this model we were able to generate a reasonable  mass splitting between the bottom and top quarks without introducing large weak isospin violation. This is possible because we followed the Sikivie et al. \cite{sikivie} scheme, according to which the technicolor interaction preserves a custodial  $SU(2)_{V}$ symmetry.

The ratio  $R \sim \frac{1}{40}$  between  the bottom and top quark masses that arises  in our model is intimately connected to the breaking of the Right-Handed symmetry. Of course, we are trading the problem of understanding the large mass splitting between  the top and bottom quarks, to the more fundamental question, of understanding  why there is a very large splitting between   the  Left-Right gauge   symmetries.
There are several points in the model that still need to be developed as the introduction of the other families,  the origin of the ETC symmetry breaking as well as of the GUT symmetry, etc... However we expect that the basic idea of the model discussed here may represent a different approach to solve this problem.
\section*{Acknowledgments}
This research was partially supported by the Conselho Nacional de Desenvolvimento
Cient\'{\i}fico e Tecnol\'ogico (CNPq) (AAN), and by Fundac\~ao de Amparo \`a
Pesquisa do Estado de S\~ao Paulo (FAPESP) (AD).

\end{document}